\documentclass[pre,showpacs,superscriptaddress]{revtex4}
\usepackage{amsfonts}
\usepackage{amsmath}
\usepackage{amssymb}
\usepackage{graphicx}

\newcommand{\beq}{\begin{equation}}
\newcommand{\eeq}{\end{equation}}
\newcommand{\be}{\begin{eqnarray}}
\newcommand{\ee}{\end{eqnarray}}
\newcommand{\bk}{{\bf k}}
\newcommand{\bq}{{\bf q}}
\newcommand{\bx}{{\bf x}}
\newcommand{\br}{{\bf r}}

\begin{document}

\title{The smooth cut-off Hierarchical Reference Theory of fluids}
\author{Alberto Parola}
\affiliation{Dipartimento di Fisica e Matematica, Universit\`a dell'Insubria, 
Via Valleggio 11, 22100 Como, Italy}
\author{Davide Pini}
\affiliation{Dipartimento di Fisica, Universit\`a degli Studi di Milano, 
Via Celoria 16, 20133 Milano, Italy}
\author{Luciano Reatto}
\affiliation{Dipartimento di Fisica, Universit\`a degli Studi di Milano, 
Via Celoria 16, 20133 Milano, Italy}



\begin{abstract}
We provide a comprehensive presentation of the Hierarchical Reference Theory (HRT) in the 
smooth cut-off formulation.
A simple and self-consistent derivation of the hierarchy of differential equations 
is supplemented by a comparison with the known sharp cut-off HRT. Then, the theory is 
applied to a hard core Yukawa fluid (HCYF): a closure, based on a mean spherical approximation
ansatz, is studied in detail and its intriguing relationship to the self consistent Ornstein-Zernike
approximation is discussed. The asymptotic properties, close to the critical point are investigated 
and compared to the renormalization group results both above and below the critical temperature. 
The HRT free energy is always a convex function
of the density, leading to flat isotherms in the two-phase region with a finite compressibility 
at coexistence. This makes HRT the sole liquid-state theory able to obtain directly fluid-fluid
phase equilibrium without resorting to the Maxwell construction.
The way the mean field free energy is modified due to the inclusion of 
density fluctuations suggests how to identify the spinodal curve. Thermodynamic properties and
correlation functions of the HCYF are investigated for three values of the inverse Yukawa range:
$z=1.8$, $z=4$ and $z=7$ where Monte Carlo simulations are available. The stability of
the liquid-vapor critical point with respect to freezing is also studied. 
\end{abstract}
\maketitle
\section{Introduction}

Our understanding of the liquid-vapor phase transition 
was boosted, in the seventies, by
the introduction in a statistical physics framework, of genuinely 
field theoretical methods: long wavelength effective actions
and Renormalization Group (RG) techniques \cite{zinn}. 
The microscopic justification of these coarse grained approaches has been the 
subject of several studies in the
eighties \cite{fisher,hub,hrt0} paving the way to quantitative investigations of the
phase diagrams of simple liquids and mixtures \cite{app}. 

In particular, the Hierachical Reference Theory of fluids 
(HRT) \cite{hrt0,adv} can be 
considered as a successful attempt at developing the RG methods in 
the framework of a liquid-state theory. HRT is based on the 
microscopic hamiltonian of the fluid, but its RG-like structure endowes it
with a number of features that are not shared by other theories of liquids:
specifically, a non-trivial description of critical behavior that incorporates
universality and scaling and, even more remarkably, a treatment of first-order
phase transitions that does not violate the convexity of the free energy required 
by thermodynamic stability, thereby yielding rigorously flat pressure 
vs. density isotherms whenever different phases coexist 
at equilibrium. The region of phase coexistence is therefore straightforwardly
obtained by HRT as the locus of diverging compressibility, and does not
have to be recovered {\it a posteriori} by enforcing thermodynamic equilibrium 
via the Maxwell construction, that is instead necessary in other liquid-state 
theories, and may turn out to be quite hard to implement, especially 
in the critical region or in the case of binary fluids.  

Notwithstanding these qualities, the original formulation of HRT leaves room
for improvement. Specifically, its most serious deficiency consists in 
the fact that it fails to predict the discontinuity of the inverse 
compressibility expected in systems with a scalar order parameter, when 
the coexistence curve is crossed at any noncritical temperature. In fact, 
HRT predicts a diverging compressibility not only inside the coexistence 
region, but also on its boundary, so that the coexistence and spinodal curves
coincide \cite{first}. Besides preventing one from recovering the correct behavior of
the compressibility at coexistence, this deficiency also forbids 
the possibility for the system to survive in a single phase inside the 
coexistence region as a metastable state --- an occurrence that would be allowed 
to states located between the coexistence and the spinodal.   
One might think of this as a minor problem, since the very
concept of a spinodal curve has little to do with equilibrium thermodynamics.
However, the instance of metastability and its breakdown via the process 
of spinodal decomposition are experimentally relevant, especially in the
context of complex fluids. 
These systems, often characterized by short range, tunable potentials and/or 
competing interactions, now offer the possibility to tailor phase diagrams 
and thermophysical properties \cite{barrat} giving rise to phase boundaries
of different topologies. Non-equilibrium states
play a major role in these systems, being rather ubiquitous in colloidal physics.
In particular, the relevance of spinodal decomposition for the occurrence of attractive
gel phases has recently been pointed out \cite{plu}. 

Here we present a new formulation of HRT, the smooth cut-off HRT, that is free from
the aforementioned shortcoming. We will detail 
the physical motivations at the basis of this generalization, 
together with the technical implementation of the smooth cut-off HRT 
to a rather flexible class of models: the Yukawa 
fluids. For this class of potentials, the smooth cut-off HRT allows also to implement exactly the
requirement that the particles be mutually impenetrable because of 
the hard-core part of the interaction, usually referred 
to as the core condition.
A numerical study of the resulting equations
allows to test the accuracy of the theory via the
comparison with available numerical simulations. 

The concept at the heart of HRT is the close relationship between the amplitude of 
density fluctuations of given wavevector $k$ and the corresponding Fourier 
component of the attractive part of the potential. 
This observation has a long history, dating back to the early works on the 
Random Phase Approximation \cite{rpa} and to the Hubbard-Schofield 
mapping of a fluid model onto a scalar field theory \cite{hub}.  
In order to make this correspondence explicit, it is customary to
split the two body inter-particle potential 
into the sum of a reference (mainly repulsive) part $v_R(r)$ and a 
residual attractive tail $w(r)$. 
The formal perturbation series in powers of $\tilde w(k)$ 
(hereafter a ``tilde" denotes Fourier transforms)
can be reinterpreted,
at all orders, as the diagrammatic expansion of a suitable effective scalar field
theory governing density fluctuations, thereby providing a bridge between 
standard liquid state theory and the
field theoretical implementation of RG \cite{adv}. 
According to the momentum space RG philosophy \cite{rg},
the order parameter fluctuations on different length-scales
are gradually introduced into the system in an iterative way, starting 
from short wavelengths (i.e. large wavevectors) down to $k=0$. 
As a consequence, the liquid-vapor phase transition, whose order parameter is just the
$k=0$ component of density fluctuations $\rho_k$, is inhibited
at any intermediate step of the RG flow and is recovered only at the end of 
the iterative procedure. This sort of ``regularization" implicit in the RG allows
the use of rather crude approximations at each step of integration, still preserving
the scaling properties of the theory. This procedure can be conveniently exported into 
liquid state theory by taking advantage of the previously outlined correspondence
between the amplitude of density fluctuations $\rho_k$ and Fourier components of the potential
$\tilde w(k)$: {\it the effects of $\tilde w(k)$
have to be introduced into the model selectively in the wavevector $k$}. Such a criterion,
however, leaves an ample range of possibilities in the practical implementation of the 
RG procedure. The most natural choice is to define a sequence of fictitious
interactions $\tilde w_Q(k)$ with vanishing Fourier components for $k<Q$. 
By following the flow of the physical properties
of the model fluid as $Q$ varies from infinity to zero, we mimic the action of the 
momentum space RG approach: 
this program has been pursued in the usual ``sharp cut-off" formulation of HRT \cite{hrt0,adv}. 
However,
in this way the two physically sensible models, the reference system (recovered for $Q=\infty$) and
the fully interacting fluid (corresponding to $Q=0$) are connected by a sequence of artificial
models, whose interaction is characterized by long range oscillatory tails due
to the sharp discontinuity of $\tilde w_Q(k)$ in momentum space. Such a pathological interaction
may pose some specific problem for the formulation of the approximations necessary to 
evaluate the physical properties of the model along the flow, i.e. at intermediate $Q$'s. 
It would be desirable to define a different sequence of intermediate systems which, 
still maintaining the selective inclusion of wavevectors, correspond to finite range, 
regular interactions at each step. A possible solution of the problem, the
smooth cut-off HRT, was already put forward several years ago \cite{ap} but 
then the analysis did not go beyond a preliminary investigation on the critical properties
predicted by this approach, later supplemented by a study of the first order transition 
\cite{ion}. In the following we will review the formulation of the
smooth cut-off HRT equations providing full details on the theory and a 
thorough discussion of the results, some of which have been already 
anticipated in Ref. \cite{prl}.

\section{The smooth cut-off HRT equations}

Following the program previously outlined, we consider a gas of classical particles 
in dimension $d$,
interacting via a spherically symmetric two body potential $v(r)$ written as the 
sum of a reference part and a tail $w(r)$:
\beq
v(r)=v_R(r)+w(r)
\label{split}
\eeq
Then we define a sequence of intermediate systems interacting with potential $v_t(r)$ 
parametrized by $t\in (0,\infty)$ which interpolates 
between $v_R(r)$ (for $t=0$) and $v(r)$ (as $t\to \infty$). This procedure is able to 
capture the RG spirit whenever
$v_t(r)$ {\it effectively suppresses the Fourier components of $w(r)$
with $k < Q\div  e^{-t}$}. This effect can be obtained by different definitions
of $v_t(r)$. Here we concentrate on the following class of parametrizations  \cite{ap}:
\be
v_t(r)&=&v_R(r) + w_t(r) \nonumber \\
&=&v_R(r) + \left [ w(r) - \psi(t)\,e^{-dt}\,w(r\,e^{-t})\right ] 
\label{vt}
\ee
where $\psi(t)$ is a monotonically decreasing 
switching function, to be specified later, 
vanishing as $t\to\infty$ and satisfying $\psi(0)=1$. 
The Fourier transform of $w_t(r)$ is given by:
\beq
\tilde w_t(k) = \tilde w(k) - \psi(t)\, \, \tilde w(k\,e^t)
\label{wt}
\eeq
For a short range interaction characterized by an attractive
tail with Fourier components $\tilde w(k)$, negligible beyond a certain wave vector $Q^*$,
the intermediate potential $\tilde w_t(k)$ coincides with $\tilde w(k)$ for $k > Q^*\,e^{-t}$ 
while $\psi(t)<1$ inhibits the Fourier components at $k<Q^*$.
Therefore the definition (\ref{vt}) does satisfy the 
requirements discussed in the introduction,
gives rise to a smooth, regular interaction at all $t$'s and may represent the starting point for 
a microscopic implementation of the RG approach. 

First order perturbation theory \cite{hansen} provides the exact expression for the 
change in the free energy density of this model when $t$ is varied:
\begin{equation}
\frac{d A_t}{dt}=-\beta\,\frac{\rho^2}{2}\int d\br \,g_t(r) \,\frac{d w_t(r)}{dt}
\label{hrt0}
\end{equation}
where $A_t$ is minus the excess free energy density divided by $k_BT$,
$g_t(r)$ is the radial distribution function of the intermediate system and 
$\beta=1/k_BT$.
This equation relates the change in the free energy with the change in the interaction
but requires the knowledge of the two body correlations. An analogous equation for the
pair correlations can be obtained starting from the formal diagrammatic expansion of the
free energy functional in powers of the interaction, as derived in Ref. \cite{adv}:
\begin{equation}
\frac{d }{dt}\,\tilde c_t^{(2)}(k)=-\beta\,\int \frac{d\bq}{(2\pi)^d} \,\left [\frac{1}{2}
\tilde c_t^{(4)}(\bk,-\bk,\bq,-\bq) + \tilde c_t^{(3)}(\bk,-\bq,\bq-\bk) \, 
\tilde F_t^{(2)}(|\bk-\bq|)\,
\tilde c_t^{(3)}(-\bk,\bq,\bk-\bq) \right ] 
\left[\tilde F_t^{(2)}(q)\right ]^2\,\frac{d \tilde w_t(q)}{dt}
\label{hrt1}
\end{equation}
where 
$\tilde c_t^{(n)}(\bk_1\dots \bk_n)$ 
is the $n$-particle direct correlation function (including the ideal gas contribution)
defined in terms of the $n^{th}$ functional density derivative of the total 
free energy of the partially interacting system \cite{adv}. In particular, 
$\tilde c_t^{(2)}(k)$ is related to the usual direct correlation function $\tilde c_t(k)$ by
$$
\tilde c_t^{(2)}(k) = -\frac{1}{\rho}+\tilde c_t(k) .
$$
The Ornstein-Zernike equation connects $\tilde c_t^{(2)}(k)$ to $\tilde F_t^{(2)}(k)$,
which equals the structure factor $S_t(k)$ multiplied by $\rho$:
$$
\tilde F_t^{(2)}(k)= \rho\,S_t(k)=
\frac{\rho}{1-\rho\,\tilde c_t(k)}   
=\rho+\rho^2 \,\int d\br\,e^{i\bk\cdot\br} \left [ g_t(r)-1\right ] .
$$ 
Equations (\ref{hrt0},\ref{hrt1}) are the first two terms of an exact 
hierarchy of differential equations 
for the free energy and the many body direct correlation functions of the model. Together with the
definition of $w_t(r)$ (\ref{vt}) this hierarchy defines the smooth cut-off formulation of HRT. 
It is interesting to notice that when Eq. (\ref{hrt1}) is evaluated at vanishing external
momentum ($k=0$), it becomes equivalent to the second density derivative of the HRT equation
for the free energy (\ref{hrt0}). An analogous correspondence takes place throughout the full
hierarchy, the $(n+1)^{th}$ differential equation being related to the density derivative of the
$n^{th}$. This coincidence is not accidental, because the direct
correlation functions are just functional derivatives of the free energy, and the functional
derivative evaluated at zero momentum reduces to the partial derivative with respect to the
uniform density. This shows that the first HRT equation (\ref{hrt0}) does contain 
information on the structure of the full hierarchy by virtue of the exact relationship between 
the free energy and the direct correlation function implied by the compressibility sum rule: 
\beq
\int \,d\br\,c_t(r) = \frac{\partial^2 A_t}{\partial \rho^2} .
\label{comp} 
\eeq
In order to exploit the connection of the HRT
equations with the RG theory, it is convenient to rescale appropriately 
the set of direct correlation functions. As previously discussed, the switching parameter 
$t$ provides the wavevector scale of density fluctuations $Q\div e^{-t}$. 
At criticality, according to general scaling arguments \cite{rg}, the expected 
behavior of the $n$-particle direct correlation function (also known as one particle irreducible
$n$-point function) is then given by 
\beq
\tilde c_t^{(n)}(\bq_1\dots\bq_n) = -e^{\left [\frac{n}{2}(d-2+\eta)-d\right ]\,t}\,
u_t^{(n)}(\bq_1e^{t}\dots\bq_ne^{t})
\label{scaling}
\eeq
where 
$\eta$ is the critical exponent related to the anomalous dimension \cite{zinn}.
The scaling form which characterizes correlations at the critical point is recovered provided 
the function $u_t^{(n)}(\bx_1\dots\bx_n)$ tends to a finite limit as $t\to\infty$. 
By substituting Eq. (\ref{scaling}) into the full (infinite) set of 
HRT equations, it is easy to show that 
the explicit dependence on the parameter $t$ disappears from the equations for the scaling functions
$u_t^{(n)}$ if the switching function is given by $\psi(t)=e^{-(2-\eta)t}$. 
With this choice of $\psi(t)$, the full hierarchy, rescaled through the definition (\ref{scaling}),
admits a fixed point solution providing the formal link with RG methods and allowing for a 
microscopic derivation of the known results on scaling laws and critical exponents \cite{ap}
in full agreement with the known dimensionality expansion \cite{zinn}. 
Note however that the scaling form is expected to hold only at small wavevectors, i.e. at 
large $t$, implying that the HRT equations provide a microscopic implementation of the RG 
procedure for any choice of $\psi(t)$ characterized by the asymptotic behavior:
\beq
\lim_{t\to\infty} e^{(2-\eta)t}\,\psi(t) = \psi_\infty
\label{asinto}
\eeq
with $\psi_\infty$ finite and non zero. 

\subsection{Sharp { vs.} smooth cut-off procedure}

The original implementation of HRT \cite{hrt0} was based on a 
different switching-on 
of the attractive
interaction, inspired by the momentum shell integration RG \cite{rg}. The  sequence of
intermediate systems are defined by an attractive part of the potential $w_Q(r)$ whose Fourier
components are set to zero below a cut-off wave vector $Q$. By varying $Q$ from infinity to zero
$v_Q(r)=v_R(r)+w_Q(r)$ interpolates between the reference system and the fully 
interacting one. In this case, however,
the ``slice" of potential $\delta w_Q\equiv w_{Q-dQ}-w_Q$ 
included at each step is characterized by a finite Fourier transform $\tilde w(Q)$ 
in an infinitesimal domain $(Q-dQ,Q)$. Such a singular form of $\delta w_Q$ has profound 
consequences on the structure of the HRT equations: now,
first order perturbation theory is not sufficient to evaluate the
change in the free energy when $Q$ is infinitesimally decreased 
and a resummation of the ring diagrams is required \cite{hrt0}. Therefore, the HRT equations 
in the sharp cut-off formulation do not coincide with Eqs. (\ref{hrt0},\ref{hrt1}): 
in particular, the momentum integrations
are limited to the shell $q=Q$ suggesting a decoupling of the long wavelength properties of
the fluid model.
In the critical region, the sharp cut-off HRT hierarchy
acquires a form independent of the interparticle interaction, leading in a natural way to the
concept of universality 
of the critical behavior. This appealing feature is not shared by the 
smooth cut-off hierarchy where the proof of universality requires a 
more elaborate analysis of the equations \cite{ap}. 
On the other hand, the intermediate potentials $w_Q(r)$ 
in the sharp cut-off formulation display unphysical long range 
oscillatory tails as a consequence of the discontinuity in Fourier space. This 
circumstance brings about some difficulty in the formulation of an approximate closure to
the infinite hierarchy of differential equations, due to the absence of an underlying 
physical intuition on the properties of fluid models characterized by such a class of potentials. 
The simple Ornstein-Zernike (OZ) closure to the first equation of the hierarchy
adopted in all the previous investigations of HRT,
although provides non classical critical exponents and scaling law together with flat 
isotherms at coexistence, has been proved to 
induce some artificial feature in the description of first order phase transitions 
leading to the divergence of the compressibility at the phase boundary, 
i.e to the coincidence of the spinodal with the binodal line \cite{first}. 
Here we will show how the same class of OZ closures instead provides a satisfactory description 
of the phase diagram of simple fluids within the smooth cut-off formulation of the 
HRT equation for the free energy (\ref{vt},\ref{hrt0},\ref{asinto}). 

\section{A simple closure for Yukawa potentials}

The OZ closure we are going to study (named HRT-OZ) 
amounts to parametrize the radial distribution function
for the intermediate system $g_t(r)$ appearing in the first equation of the HRT hierarchy 
(\ref{hrt0}). Here we will focus on a simple yet flexible class of interactions: the 
hard core plus Yukawa potentials defined as the sum of a pure hard core term of diameter 
$\sigma$ and an attractive Yukawa tail of inverse range $z$
\beq
w(r)=-\epsilon\,\frac{e^{-z(r-\sigma)}}{r}
\label{yuk}
\eeq
In the following, $\sigma$ and $\epsilon/k_B$ will be taken as units of length and temperature
respectively (i.e. we set $\sigma=1$, $\epsilon=k_B$). 

As previously noticed, in order to ensure that a memory of
the full structure of the HRT hierarchy is contained in the free energy equation (\ref{hrt0}),
the compressibility sum rule (\ref{comp}) has to be satisfied within the parametrization.
A simple way to implement such a requirement is to adopt an approximate form of the 
correlation functions coming from some accurate liquid state theory suitably generalized in 
order to allow for the consistency between free energy and correlations (\ref{comp}).
In order to take advantage of the analytical form available for the correlation functions
of a Yukawa fluid, we have chosen to use, as a supporting liquid state theory, the Mean Spherical 
Approximation (MSA) defined by the usual core condition ($g_t(r)=0$ for $r<1$) supplemented by 
the explicit form of the direct correlation function outside the core:
\begin{eqnarray}
c_t(r) &=& -\beta \left ( w_t(r) +\lambda_t \,w(r) \right ) \nonumber \\
&=& -\beta (1+\lambda_t)\,w(r) +\beta \psi(t)\,e^{-dt}\,w(r\, e^{-t})  \qquad \qquad {\rm for} \quad r>1
\label{closc}
\end{eqnarray}
where $\lambda_t$ is implicitly defined by the compressibility sum rule (\ref{hrt0}) in terms 
of the excess free energy $A_t$. 
The radial distribution function and the direct correlation function are connected via
the OZ equation. 
Within MSA the direct correlation function is always analytic in $k^2$ at small $k$, even at 
the critical point, implying that the critical exponent $\eta$ vanishes within this
class of approximations.
For the HCYF, Eq. (\ref{closc}) implies a 
direct correlation function written as the sum of {\it two} Yukawa's of different 
ranges $z_i$ and amplitudes $K_i$:
\beq
c_t(r)= K_1 \frac{e^{-z_1(r-1)}}{r}+K_2 \frac{e^{-z_2(r-1)}}{r}
\label{yuk2}
\eeq
outside the particle core (for i.e. $r>1$), with
\be
z_1 &=& z \nonumber \\
z_2 &=& z\,e^{-t} 
\label{parz}
\\
K_1 &=& \frac{1+\lambda_t}{T}  \nonumber \\
K_2 &=& -\frac{e^{z(1-e^{-t})}}{T}\,\psi(t)\,e^{-(d-1)t}
\label{park}
\ee
In three dimensions, the MSA integral equation
can be solved analytically for this class of parametrizations 
and the solution reduces to a set of algebraic equations \cite{hoye}. 
Within HRT-OZ, this set 
is coupled to the evolution equation (\ref{hrt0}) via the compressibility 
sum rule (\ref{comp}) which implicitly defines the parameter $\lambda_t$ present in $K_1$.
The explicit form of the equations, together with a detailed description of the
numerical scheme adopted for their resolution, can be found in Appendix A. 
Here we remark that, as discussed in detail in Appendix A, the parametrization (\ref{closc})
also allows to improve the performance of the MSA (and then of the HRT-OZ) 
in the high density region, where 
the hard sphere direct correlation function is known to display a non negligible tail outside the
particle core. If this additional contribution to $c_t(r)$ is represented as a 
Yukawa tail {\it of the same inverse range $z$ as that of the attractive part of the 
potential}, its presence is equivalent to a redefinition of the parameter $\lambda_t$ for 
$t=0$, whose value is determined by Eq. (\ref{comp}) if the 
Carnahan-Starling expression for the hard sphere free energy is used. 

\subsection{HRT-OZ { vs.} SCOZA}

The Self consistent Ornstein Zernike Approximation (SCOZA) is a well known
liquid state theory which proved successful in the determination of phase diagrams in 
Yukawa fluids \cite{scoza}. Originally, SCOZA has been formulated as a generalization
of the MSA expression for the direct correlation function outside the core:
\begin{equation}
c(r)= -\beta \lambda\,w(r)
\label{scoza}
\end{equation}
where the parameter $\lambda$ is determined by requiring thermodynamic consistency 
between the internal energy and the compressibility routes to thermodynamics.
In terms of the excess free energy density (divided by $-k_BT$) $A$, this condition reads:
\begin{eqnarray}
\frac{\partial A}{\partial \beta} &=& -\, \frac{\rho^2}{2} \int d\br \,g(r)\, w(r) \nonumber \\
\frac{\partial^2 A}{\partial \rho^2} &=& \int d\br \,c(r)
\label{scozac}
\end{eqnarray}
The two exact expressions (\ref{scozac}), together with the core condition, $g(r)=0$ for $r<1$,
and the parametrization (\ref{scoza}) for $r>1$, give rise to a partial differential 
equation for the free energy as a function of $\rho$ and $\beta$. 
The close similarity between Eqs. (\ref{scozac}) and (\ref{hrt0},\ref{comp}) suggests that 
SCOZA can be somehow related to the smooth cut-off formulation of HRT. 
In fact, by defining the special sequence of intermediate potentials $w_t(r)$ as:
\begin{equation}
w_t(r)=t\,w(r)
\label{scozaw}
\end{equation}
the HRT-OZ equations (\ref{hrt0},\ref{comp}) exactly reproduce SCOZA, showing that 
the two theories differ just by the switching-on procedure of the attractive pat of
the interaction: while SCOZA tunes the amplitude of $w(r)$, HRT changes 
simultaneously both amplitude and range of the potential. The long range 
repulsive tail introduced by HRT during the integration, 
allows to inhibit phase separation at all finite $t$'s
thereby mimicking the RG approach. 

\begin{figure}
\includegraphics[width=6cm,angle=0]{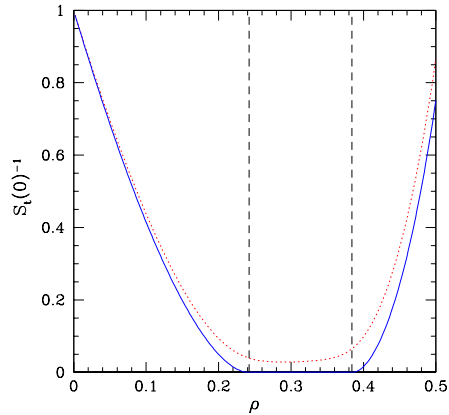}
\caption{Dimensionless inverse compressibility $S_t(0)^{-1}$ of a $z=1.8$ Yukawa fluid 
within MSA below the critical temperature ($T=1$). 
The potential is given by (\ref{vt}) with $\psi(t)=(\cosh t)^{-2}$ (see Eq. \ref{second})
at $t=2$ (dotted line) and $t=10$ (full line). 
Vertical dashed lines bracket the coexistence region: here the $t=10$ isotherm is smaller than
$10^{-7}$. 
The standard MSA result for a single Yukawa coincides with the full line outside coexistence
while no solution exists inside the coexistence region.
}
\label{press}
\end{figure}

In order to emphasize the key role played 
by the weak long range repulsive tail added to the physical interaction in the
smooth cut-off HRT, let us consider the simple MSA for 
a Yukawa fluid, which, at low temperatures, 
is known to display a region inside the spinodal where no solution exists.
However, as shown in Fig. \ref{press}, if the physical attractive Yukawa potential is supplemented by 
a long range repulsive Yukawa contribution of the form (\ref{vt}), the spinodal 
curve disappears and a solution of the MSA equation exists in the whole phase diagram.
When $t\to\infty$, the repulsive term in the potential becomes vanishingly small:  
the isotherm becomes flat and the compressibility diverges in the whole 
region inside the spinodal of the single Yukawa, thereby enforcing the convexity of the free energy.
Therefore, cutting off long wave length density fluctuations by the addition of a small, long ranged 
repulsive tail appears to be 
an easy way to implement one of the most important features of equilibrium statistical 
mechanics within an approximate liquid state theory. 
Note however that within MSA, this kind of ``regularization" of the physical 
two body interaction leads to the collapse of the spinodal and coexistence curve. 

On the basis of these observations, the smooth cut-off HRT-OZ approach is then expected to
provide a convex free energy: in the following Section we will show that HRT-OZ enjoys other 
relevant physical properties of first and second order phase transitions. 

\section{Asymptotic analysis}

As already noticed, the critical properties of the theory are determined by long wavelength 
fluctuations which come into play at the last stages of the $t$-evolution, i.e. in the 
limit $t\to\infty$. The full HRT-OZ equations, in the critical region and at large $t$'s 
considerably simplify: remarkably, the formal structure of the
asymptotic HRT-OZ equation is in fact identical to that obtained by a much simpler closure to the 
first equation of the hierarchy (\ref{hrt0}) based on the Random Phase Approximation (RPA), i.e. 
neglecting the core condition. In fact, according to the critical phenomena paradigm, short range
correlations should not affect the physics at long wavelengths which dominates in the critical region. 
And this property is explicitly satisfied within HRT-OZ. 

\begin{figure}
\includegraphics[width=8cm,angle=0]{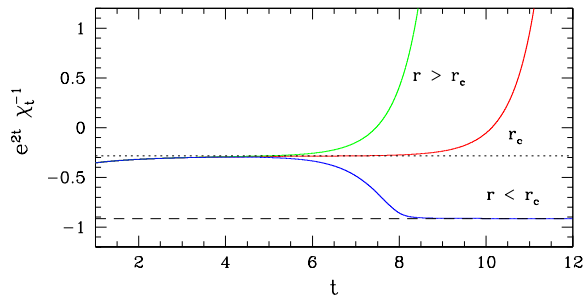}
\caption{Evolution of the scaled inverse compressibility (\ref{defx2}) on 
the critical isochore $\varphi=0$ for 
$p=0.5$, $u=0.1$ and three temperatures: above ($r=-0.30810$), below ($r=-0.30811$)
and very close to the critical temperature ($r=-03081037$).
obtained by numerical integration of Eq. (\ref{hrt6b}). The dotted line is the
critical fixed point ($x^*=-0.2848\dots$) obtained by standard RG analysis of Eq. (\ref{fix2}).
The dashed line is the low temperature fixed point $x_<=p-2\sqrt{p}=-0.9142\dots$ (see 
Ref. \cite{ion}).
}
\label{evol}
\end{figure}

In the simplified RPA form, the direct correlation function reads
\begin{equation}
c_t(r) = c_R(r) -\beta \left ( w_t(r) +\lambda_t \,w(r) \right )
\label{closcrpa}
\end{equation}
for all $r$'s. As usual, the parameter $\lambda_t$ is 
determined by the compressibility sum rule (\ref{comp}). 
Appendix B contains the details of the asymptotic analysis of (\ref{hrt0}) under both 
closures (\ref{closcrpa}) and (\ref{closc}). Here we simply discuss the physical contents of the
result. It is convenient to first define a modified free energy density (in units of $k_BT$)
${\cal A}_t$ which encompasses both the ideal gas term ($A^{id}$) and the 
mean field contribution due to that part of the interaction not already included in $A_t$:
\beq
{\cal A}_t = A_t + A^{id}-\beta\, 
\frac{\rho^2}{2} \left [ \tilde w(0)-\tilde w_t(0)\right ]
\label{moda}
\eeq
The evolution equation (\ref{hrt0}) under the closure (\ref{closcrpa}) 
for a three dimensional Yukawa fluid in the $t\to\infty$ limit 
and in the critical region becomes (see Appendix B): 
\begin{equation}
\frac{\partial {\cal A}_t}{\partial t} = e^{-3t}\,
\frac{z^3}{4\pi} \left [ 1+ p^{-1/2} \frac{2-p -\sqrt{p}\, \sqrt{1+x}}
{\sqrt{p+x+2\sqrt{p}\, \sqrt{1+x}}}\right ]
\label{hrt6}
\end{equation}
where 
\beq
x=-\frac{z^2}{4\pi\beta\psi_\infty} \,e^{2t}\, \frac{\partial^2{\cal A}_t}{\partial \rho^2}
\label{defx}
\eeq
is a renormalized inverse compressibility 
and the non universal parameter $p<1$ is the ratio between
the curvature of the potential and of the direct correlation function divided by $\psi_\infty$ 
(\ref{asinto}), as shown in Eqs. (\ref{prpa},\ref{pmsa}). 
It is convenient to remove an analytic contribution and a 
regular pre-factor in Eq. (\ref{hrt6}) by a suitable rescaling of 
the free energy ${\cal A}_t \to\Psi_t$ and of the density $\rho\to\varphi$ as detailed 
in Eq. (\ref{rescale}):
\beq
\rho -\rho_c = \varphi\,\left ( \frac{z^{5}}{(4\pi)^2\beta\sqrt{p}} \right )^{1/2} .
\eeq
The resulting evolution equation reads:
\beq
\frac{\partial\Psi_t}{\partial t} = -e^{-3t}\,U(x)=-e^{-3t}\,
\frac{2-p -\sqrt{p}\, \sqrt{1+x}} {\sqrt{p+x+2\sqrt{p}\, \sqrt{1+x}}}
\label{hrt6b}
\eeq
where (\ref{defx}) is now expressed as
\beq
x = e^{2t}\, \frac{\partial^2\Psi_t}{\partial \varphi^2}
\label{defx2}
\eeq
Notice that, according to the definition (\ref{defx}), the variable $x$ 
may attain a finite limit ($x^*$) for $t\to\infty$ only if the compressibility diverges,
meaning that the fully interacting system is either at the critical point or in the two phase
region. The asymptotic Eq. (\ref{hrt6b}) has been numerically integrated for 
an initial condition at $t=0$ 
of the Landau-Ginzburg type, i.e. appropriate for a self-interacting scalar field theory:
\beq
\Psi_0(\varphi) = \Psi_0(0) + r\,\varphi^2 + u\,\varphi^4
\label{phi4}
\eeq
Besides $p$, which retains information on the range of the potential,
this effective model is defined by two parameters: $r$ is a 
measure of temperature (shifted by the mean field critical value) while $u>0$ is assumed to 
be state independent. 
The typical behavior of the rescaled inverse compressibility $x(\varphi,r)$ during the ``time" 
evolution determined by Eq. (\ref{hrt6b}) is shown in Fig. \ref{evol} for $p=0.5$, $u=0.1$ 
and three states on the critical isochore $\varphi=0$: above, below and very close to 
the critical temperature $r_c$. 
Above $r_c$ the divergence of $x$ shows that the inverse compressibility attains a finite, positive
value, as expected in supercritical fluids. Close but still above the critical point, 
the evolution of $x$ displays a plateau at $x^*<0$ before diverging again. 
Instead, for $r<r_c$ the evolution crosses $x^*$ and asymptotically sets to a finite, negative value 
$x_<$ implying that the inverse compressibility vanishes. 
Inside the coexistence curve the fluid is 
therefore characterized by flat isotherms, i.e. the Maxwell construction is recovered via the
inclusion of long wavelength fluctuations. Such a behavior of the ``time" evolution
suggests the possible presence of two distinct fixed points of the
evolution equation, $x^*$ and $x_<$, governing
the critical and subcritical region respectively (while the ``high temperature fixed point" is
located at $x=+\infty$). 

\subsection{Critical behavior}

\begin{figure}
\includegraphics[width=12cm,angle=0]{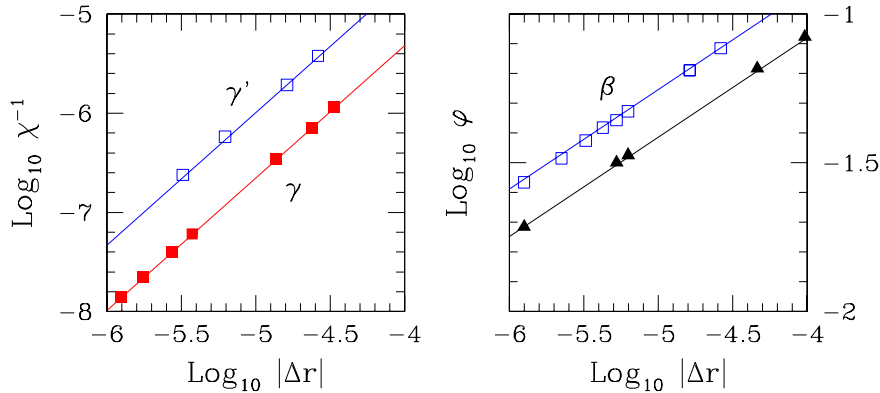}
\caption{Left panel: Log-Log plot of the inverse compressibility vs. reduced temperature for
$p=0.5$ and $u=0.1$.
Upper points are the HRT-OZ results below the critical temperature in the homogeneous 
phase, extrapolated to the coexistence curve. Lower points are on the critical isochore
$\varphi=0$ above $r_c$. Right panel: Log-Log plot of the coexistence 
curve (squares) and of the spinodal (triangles) for $p=0.5$ and $u=0.1$. 
Lines mark the asymptotic power law behavior with the critical
exponents quoted in Table I: $\gamma=1.335$ and $\beta=0.334$.
}
\label{exp}
\end{figure}

The singular behavior of the thermodynamic properties in the critical region can be 
obtained from the asymptotic form of the evolution equation (\ref{hrt6b}) following the fixed
point analysis developed in the RG theory. As a first step we eliminate from Eq. (\ref{hrt6b})
the explicit dependence on the evolution parameter $t$ by 
a $t$-dependent rescaling of the density and free energy density: 
\be
\zeta &\equiv& e^{\frac{t}{2}}\,\varphi \nonumber \\
\Gamma_t(\zeta) &\equiv& \left [\Psi_t (\varphi) -\Psi_t(0)\right ] \,e^{3t}
\label{fix1}
\ee
Notice that the rescaled inverse compressibility
$x$ (\ref{defx}) is simply expressed in terms of $\Gamma_t$ by 
$x=\frac{\partial^2 \Gamma_t}{\partial\zeta^2}$.
The evolution equation for the ``renormalized" quantities now acquires a typical RG structure: 
\beq
\left [\frac{\partial}{\partial t} + \frac{1}{2}\,\zeta\,\frac{\partial}{\partial\zeta}-3 \right ] 
\,\Gamma_t(\zeta) = 
U\left(\frac{\partial^2 \Gamma_t}{\partial\zeta^2}\right)_{\zeta=0} -
U\left(\frac{\partial^2 \Gamma_t}{\partial\zeta^2}\right)  
\label{fix2}
\eeq
where $U(x)$ is given by Eq. (\ref{hrt6b}). 
Equation (\ref{fix2}) admits a non trivial fixed point:  
i.e. a solution $\Gamma^*(\zeta)$ independent of
$t$, regular and not identically zero on the whole real axis $\zeta\in (-\infty,\infty)$.  
A fixed point corresponds to a diverging compressibility at $\varphi=0$, leading to
$x^*=\frac{d^2 \Gamma^*}{d\zeta^2}\vert_{\zeta=0}\sim -0.2848\dots$ (for $p=0.5$), 
which perfectly agrees with 
the position of the plateau in the evolution shown in Fig. \ref{evol} for $r\sim r_c$, 
and therefore identifies the critical point. Linearizing the evolution equation 
(\ref{fix2}) around $\Gamma^*(\zeta)$ we can investigate the stability properties of the fixed 
point: letting $\Gamma_t(\zeta)=\Gamma^*(\zeta)+\delta \Gamma_t(\zeta)$ 
and keeping terms up to first order in 
$\delta \Gamma_t$ we find solutions of the form 
$\delta \Gamma_t(\zeta)=e^{\Lambda t} \,\gamma(\zeta)$ where 
$\Lambda$ is the eigenvalue and $\gamma(\zeta)$ is the corresponding eigenfunction. 
As usual in this context we find two unstable eigenmodes, 
characterized by positive eigenvalues \cite{adv} governing displacements from the 
critical point along the critical isotherm and isochore.  
The first is odd in $\zeta$ and can be expressed as $\gamma(\zeta)=\frac{d \Gamma^*}{d\zeta}$ 
with $\Lambda_{o} = 1/2$, while the other, even in $\zeta$, must be computed numerically. 
The RG structure of the HRT-OZ evolution equation (\ref{fix2}) 
shows that scaling and hyperscaling are correctly reproduced by the theory.
Due to the known relation between eigenvalues and critical exponents \cite{adv}
\be
\delta &=& 1+\frac{2}{\Lambda_{o}} \nonumber \\
\gamma &=& \frac{2}{\Lambda_{e}} 
\ee 
we find $\delta=5$, as implied by scaling when the correlation exponent $\eta$ vanishes, 
while the other critical exponents follow from the numerical determination 
of $\Lambda_{e}$ and are shown in Table I for few choices of the parameter $p$: 
HRT-OZ displays non classical critical exponents weakly dependent on $p$.
It is in fact apparent that, contrary to the sharp cut-off HRT-OZ approach, here the asymptotic
equation (\ref{hrt6b}) does not acquire a universal form but retains memory of the microscopic
model through the value of the parameter $p$. Therefore the 
critical properties of the three dimensional fluid within our approximation will generally violate
universality and both critical exponents and scaling laws will depend on the precise value of $p$.
This situation closely resembles the case of non perturbative approximations 
to the RG flow equations,
whose results do depend on the cut-off function \cite{wetterich}. Note however that, as shown in
Ref. \cite{ap,ion}, the critical exponents are correct to first order in a $\epsilon=4-d$ 
expansion of the HRT-OZ equation (\ref{hrt0}) and, 
within a systematic $\epsilon-$expansion of the full HRT hierarchy, 
the dependence on $p$ disappears and universality is indeed recovered. 
\begin{table}
\vskip 0.2cm
\begin{tabular}{|c||c|c|c|c|c|c|}
\hline
Exponent & $\alpha$   & $\beta$   & $\gamma$   & $\delta$    & $\eta$ & $U_2=C_+/C_-$ \\
\hline
``Exact"    & 0.110 & 0.327 & 1.237 &   4.789 & 0.036 & 4.76  \\
\hline
HRT ($p=0.25)$ & 0.007 & 0.332 & 1.329 & 5 & 0 &   4.2 \\
\hline
HRT ($p=0.50)$ & -0.003 & 0.334 & 1.335 & 5 & 0 &  4.5  \\
\hline
HRT ($p=0.75)$ & -0.010 & 0.335 & 1.340 & 5 & 0 &  4.6  \\
\hline
\end{tabular}
\caption{HRT-OZ estimates of the critical exponents and compressibility
amplitude ratio in three dimensions for few values of the parameter $p<1$ 
compared to the exact values \cite{field} obtained by extrapolation
of high-temperature series expansions. }
\end{table}
The critical exponents can be also evaluated by numerical integration of the asymptotic
evolution equation (\ref{hrt6b}) close to the critical point with no
reference to any RG procedure. The results, shown in Fig. \ref{exp} 
for the case $p=0.5$ and $u=0.1$ conform to the expectation of the 
fixed point analysis previously outlined both above and below the
critical temperature. 

The renormalized equation (\ref{fix2}) also allows to determine 
other relevant properties of the critical regime. In Table I we show the 
compressibility amplitude ratio defined as $U_2=C_+/C_-$ where 
\beq
\chi (r,\varphi(r)) = C_\pm \, \left\vert \Delta r \right\vert^{-\gamma}
\eeq
$\chi$ is the isothermal compressibility and $\Delta r=r-r_c$.
The $+$ sign corresponds to $r>r_c$ and $\varphi(r)=0$, while the $-$ sign refers to
$r<r_c$ and $\varphi(r)$ along the coexistence curve. 
Note that in the numerical solution of the evolution equation,
the value of the inverse compressibility on the binodal, i.e on the point of
discontinuity, is not directly available and must be obtained by extrapolation of
the results at neighboring mesh points. This limits the possibility to get arbitrarily
close to the critical point and to obtain accurate estimates of the amplitude ratio $U_2$.

\begin{figure}
\includegraphics[width=8cm,angle=0]{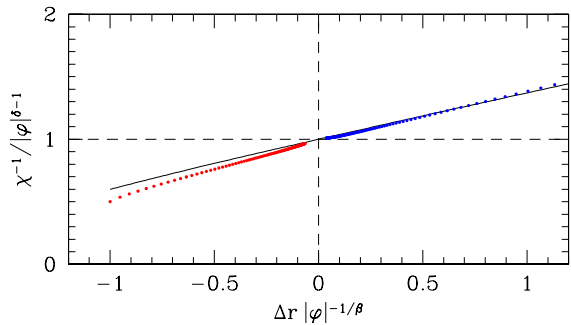}
\caption{Scaling form of the equation of state $W(x)$ as defined in Eq. (\ref{eos}) for 
$p=0.5$. Symbols: HRT-OZ results
obtained by numerical integration of Eq. (\ref{hrt6b}). 
Full line: polynomial approximant of the 
exact scaling function \cite{field}. 
The metric factors fixing the units of $r-r_c$ and $\varphi$
are determined so as to locate the coexistence curve at $\Delta r=-|\varphi|^{1/\beta}$ and 
$\chi^{-1}=|\varphi|^{\delta-1}$ at $r_c$.
}
\label{scala}
\end{figure}

The scaling form of the equation of 
state of the model can be also obtained in the form: 
\be
\chi^{-1} = \vert \varphi\vert^{\delta-1} \, W
\left (\frac{\Delta r}{\vert\varphi\vert^{1/\beta}}\right )
\label{eos}
\ee
The numerical results for $p=0.5$ 
are shown in Fig. \ref{scala} together with a polynomial approximation 
of the ``exact" scaling function from Ref. \cite{field}. Other choices of $p$ do not
modify appreciably the good agreement between HRT-OZ and RG results. 

\subsection{Below the critical temperature}

As previously noticed (see also Ref. \cite{ion}), within the smooth cut-off
HRT-OZ scheme, the inverse compressibility 
develops a singularity for $t\to\infty$ at any $r<r_c$: it attains a finite limit 
on the binodal while vanishes identically inside the two phase region. 
The way $\chi^{-1}$ drops to zero below $r_c$ is 
governed by a low temperature fixed point: 
$$
x_<=\lim_{t\to\infty} e^{2t}\frac{\partial^2\Psi}{\partial \varphi^2}
$$ 
defined by the divergence of $U(x)$ in Eq. (\ref{hrt6b}) and explicitly given by $x_<=p-2\sqrt{2}$.
The agreement between the formal analysis and the numerical results shown in Fig. \ref{evol} 
fully confirms this interpretation.

\begin{figure}
\includegraphics[width=9cm,angle=0]{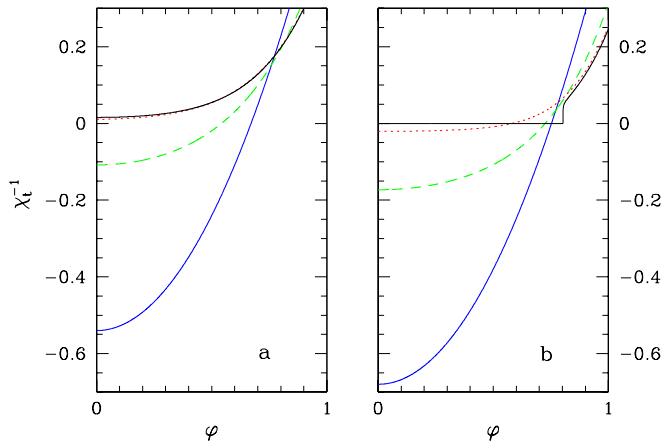}
\caption{Snapshots of the inverse compressibility as a function of density 
at four values of the evolution parameter. From bottom to top: $t=0$, $t=0.5$, 
$t=1.75$ and $t\to\infty$. Panel $a$ refers to $r=-0.27$ above the critical temperature $r_c$,
panel $b$ to $r=-0.34$ below $r_c$.
}
\label{below}
\end{figure}

\begin{figure}
\includegraphics[width=6cm,angle=0]{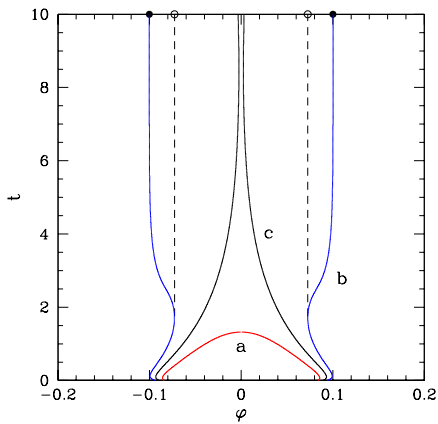}
\caption{Boundaries of the region defined by a negative compressibility as a function 
of the evolution parameter $t$ for $p=0.5$, $u=0.1$. Curve a refers to
$r=-0.27 >r_c$, curve b corresponds to  $r=-0.34 < r_c$ and curve c to 
$r=-0.308105 \lesssim_c$. 
Full dots mark the coexistence region, empty dots the spinodal. The dashed lines
are guides to the eye. The width of the unstable region has a minimum at the
near-critical temperature at $t=8.5$ and $\varphi=0.0024$.
}
\label{st}
\end{figure}

It is interesting
to examine more closely the HRT-OZ evolution of the inverse compressibility below the critical
temperature in order to understand how the inclusion of long wavelength fluctuations
drives such a 
behavior, which perfectly matches the expectations for the case of a scalar order parameter. 
In Figs. \ref{below}a,\ref{below}b we contrast the evolution of two 
typical isotherms above and below $r_c$ by showing the inverse compressibility for four values of the
switching parameter $t$, as obtained by the numerical integration of Eq. (\ref{hrt6b}). 
At $t=0$ the initial condition is a smooth function of the density which reproduces 
the Landau-Ginzburg mean field structure. Below the mean field critical temperature ($r=0$) 
it displays 
a large unstable region, corresponding to the presence of a Van der Waals loop. 
In general, the inclusion of density fluctuations, governed by the parameter $t$,
leads to a gradual increase of 
the inverse compressibility thereby reducing the width of the unstable region.
Above the true critical temperature this eventually leads to the complete suppression of
the instability (see Fig. \ref{below}a). Instead, below $r_c$, long wavelength 
fluctuations severely affect the shape of the isotherms:
$i)$ $\chi^{-1}$, when negative, flattens as a function of density and approaches zero at large $t$;
$ii)$ The isotherm sharpens at the boundaries of the interval where 
$\chi^{-1}<0$ developing a discontinuity, which defines the binodal;
$iii)$ At large length-scales the coexistence region widens again and 
in a small density interval close to the binodal the compressibility becomes positive before 
dropping to zero. 
The origin of the flattening and sharpening of the HRT-OZ isotherm has been analyzed in detail 
in Ref. \cite{ion} where the smooth cut-off HRT-OZ has been studied for a pure $\Phi^4$ field theory.
As anticipated, the smooth cut-off prescription overcomes one of the main problems
of the previous implementations of HRT-OZ \cite{first}, providing finite compressibility at coexistence. 
In this framework, we may investigate
whether it is possible to locate the boundary of the metastable phases inside the coexistence curve.  
The spinodal curve is not rigorously defined within equilibrium statistical mechanics, the 
free energy being a convex function in the whole phase diagram,  
but a special feature of the evolution shown in Fig. \ref{below}b suggests a way to 
discriminate between instability and metastability. In a metastable state we expect that 
density fluctuations on a large but finite range do not drive the system towards phase 
separation, i.e. $\chi^{-1}$ remains positive until $t$ 
does not exceed a (large) crossover value $t_s$. Only when 
density fluctuations on a macroscopic scale are included, does the first order transition 
take place. In contrast, in the truly unstable region $\chi^{-1}$ the inverse compressibility
is negative throughout the evolution. 
The curves dividing the regions of positive and negative $\chi^{-1}$ 
in the $(\varphi,t)$ plane at different temperatures are shown in Fig. \ref{st}, together with
the extent of the metastable region obtained via the criterion previously outlined. 
According to this definition, the shape of the spinodal line in a $\varphi$ vs. $\Delta r$ 
diagram appears to be characterized by the same critical exponent $\beta$ of the coexistence 
curve, as shown in Fig \ref{exp}. 
The value $t^*$ of the evolution parameter $t$ where the width of the unstable region has a minimum,
diverges when the critical point is approached. This corresponds to the divergence of a characteristic
length-scale $R_c \div e^{t^*}$, which may be identified as the critical droplet radius on the
spinodal line, according to the standard droplet model picture of nucleation \cite{drop}. 
This divergence follows a power law consistent with the scaling 
$R_c\div \xi \div |\Delta r|^{-\nu}$ with $\nu=\gamma/2$ (see Table I).  

\section{Numerical results and phase diagrams}

Here we show some results obtained by the numerical integration of the full smooth cut-off 
HRT-OZ equation as derived in Appendix A (\ref{pde}). We systematically investigated the 
HCYF at three values of the inverse range parameter $z$: $z=1.8$, which mimics the usual
Lennard-Jones potential, $z=4$ and $z=7$ which are more appropriate for modeling colloidal 
suspensions. The numerical solution of the partial differential equation has 
been performed by a full implicit predictor-corrector
finite difference method on a density mesh of $2000$ points. The estimated numerical error 
cannot be appreciated on the scale of the figures. 

\subsection{Spinodal and binodal}

\begin{figure}
\includegraphics[height=6cm,angle=0]{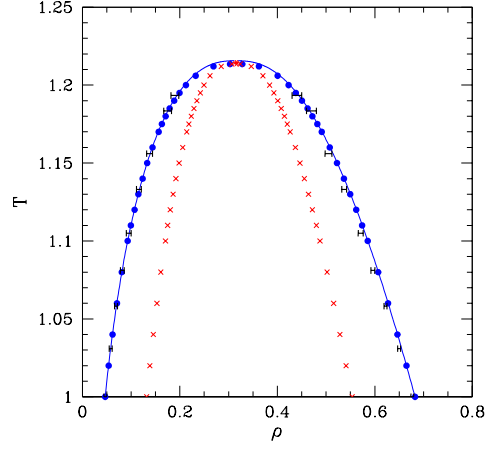}
\caption{Coexistence curve and spinodal of the HCYF for $z=1.8$: HRT-OZ (full circles),
SCOZA (line) and Monte Carlo \cite{scoza} (error bars). The HRT-OZ spinodal is also shown 
(crosses). 
} 
\label{coexz18}
\end{figure}

\begin{figure}
\includegraphics[height=6cm,angle=0]{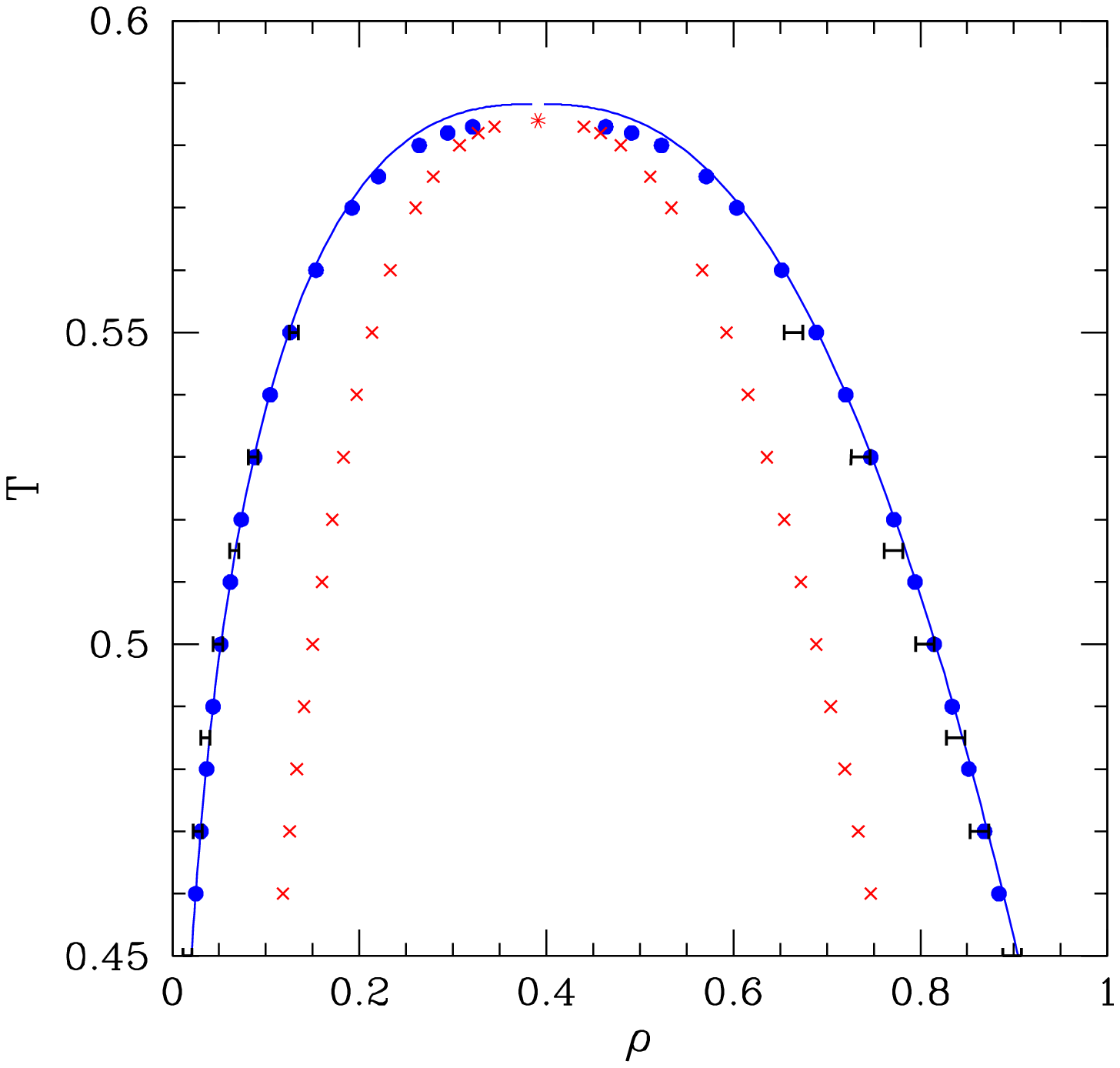}
\caption{Coexistence curve and spinodal of the HCYF for $z=4$. Notation as in Fig. \ref{coexz18}. 
The Monte Carlo simulations are taken from Ref. \cite{duda}. 
} 
\label{coexz4}
\end{figure}

\begin{figure}
\includegraphics[height=6cm,angle=0]{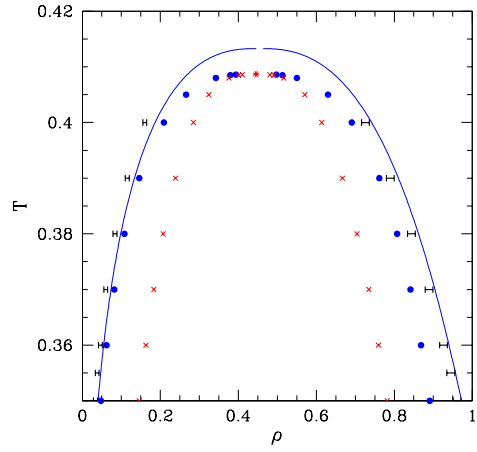}
\caption{Coexistence curve and spinodal of the HCYF for $z=7$. Notation as in Fig. \ref{coexz18}.
The Monte Carlo simulations are taken from Ref. \cite{duda}.
}
\label{coexz7}
\end{figure}

The coexistence curve can be immediately read-off the numerical solution of the equation, for
in the two phase region the inverse compressibility vanishes. The spinodal curve instead is 
evaluated following the criterion discussed in Section (IV-B) and illustrated in Fig. \ref{st}. 
The results are shown in Figs. \ref{coexz18},\ref{coexz4},\ref{coexz7} together with the 
coexistence curves obtained by Monte Carlo simulations and SCOZA. 

The HRT-OZ coexistence curve agrees very well with the Monte Carlo data for $z=1.8$ and $z=4$
while some deviation appears at $z=7$ where the MSA closure, which underlies the present 
implementation of the smooth cut-off HRT-OZ, is less accurate. As expected, the coexistence curve
becomes flatter by reducing the range of the attractive interaction, as testified by 
the different scale of the temperature axis in the three cases. Note that the SCOZA
results are always close to the HRT-OZ binodal, with some deviation in the critical region where 
the HRT-OZ coexistence line is flatter. Again, the largest discrepancy between
HRT-OZ and SCOZA occurs at high $z$, however, even in this case, 
the two critical temperatures differ by less than $1\%$.

\subsection{The liquid-solid phase boundary}

The fluid-solid phase equilibrium of the HCYF has been widely 
studied, mainly because it allows one to describe the qualitative change 
in the shape of the phase diagram brought about by changing the attraction 
range. As is well known, as the range is narrowed, i.e., as  $z$ is increased, 
one goes from the situation where the phase diagram has both a stable 
fluid-fluid critical point and a fluid-fluid-solid triple point, to that
where the critical point moves into the fluid-solid coexistence 
region, thereby making the fluid-fluid transition metastable with respect to
freezing, and there is no triple point~\cite{hagen,lomba}. The former case 
is found in atomic and molecular fluids, while the latter is characteristic 
of many colloidal 
systems, including protein solutions~\cite{lomakin}. The disappearance of 
a stable fluid-fluid transition is in fact a general feature of short-ranged 
attractive interactions, but the Yukawa potential was, together with 
the Asakura-Oosawa one~\cite{gast}, among the first to be considered 
in this respect.  
 
HRT is not a theory of freezing, but of course it can be used to predict the
equilibrium fluid-solid phase diagram, if the free energy of the solid is 
provided by some different procedure. We have resorted to the simplest and 
most adopted one~\cite{weis,hecht,gast,hagen,dijkstra}, based 
on Barker and Henderson perturbation 
theory~\cite{barker}. This approach was originally developed for the fluid
phase, but it can be straightforwardly employed in the solid one as well.  
According to it, the Helmholtz free energy 
of a solid of particles interacting via a hard-sphere plus tail potential 
is given by:
\begin{equation}
A_{s}=A_{s}^{R}-\frac{1}{2}\beta\rho^{2}
\int \!\!d{\bf r} \, g_{s}^{R}(r) \, w(r) + 
A_{2}+{\cal O}(\beta^{3})\, ,
\label{solid}
\end{equation}       
where $A_{s}$ is minus the excess free energy per unit volume of the solid 
divided by $k_{B}T$, $A_{s}^{R}$ is the corresponding quantity for the 
reference hard-sphere solid, $g_{s}^{R}(r)$ is the radial distribution 
function of the hard-sphere solid averaged over the solid angle, and $w(r)$ 
is the potential tail, i.e., the attractive Yukawa in the present case. 
Following previous investigations~\cite{gast,dijkstra}, 
Eq.~(\ref{solid}) has been truncated at the second-order term $A_{2}$,
which has been estimated by the so-called ``macroscopic-compressibility'' 
approximation, also suggested by Barker and Henderson~\cite{barker}: 
\begin{equation}
A_{2}\simeq
\frac{1}{4}\beta^{2}\rho^{2}\chi_{s}^{R}
\int \!\! d{\bf r}\, g_{s}^{R}(r)\, w^{2}(r) \, ,
\label{mc}
\end{equation}
where $\chi_{s}^{R}$ is the isothermal compressibility of the hard-sphere 
solid divided by the ideal-gas value. Comparison  
with Monte Carlo simulations on depletion potentials~\cite{rotenberg} 
has shown that, although 
the estimate of $A_{2}$ given by Eq.~(\ref{mc}) deviates quite 
substantially from its exact value, which would involve knowledge of the
three- and four-particle distribution function of the reference solid,   
the sum of $A_{1}$ and $A_{2}$ given by Eqs.~(\ref{solid},\ref{mc}) is nevertheless
remarkably close to the simulation results for the full free energy $A_{s}$,  
unless the interaction is very short-ranged.  
In fact, the truncated expansion~(\ref{solid}) has proved to be 
more accurate in the solid than in the fluid~\cite{dijkstra}, of course
provided the solid in study has the same lattice structure as the reference 
one. 

Equations~(\ref{solid},\ref{mc}) were 
already used together with the sharp cut-off version of HRT-OZ to study 
the fluid-solid equilibrium of non-additive hard-sphere mixtures, including
the Asakura-Oosawa case~\cite{nonadditive}. They were also applied together
with SCOZA to both the HCYF considered here~\cite{foffi} 
and a two-Yukawa fluid with competing attractive and repulsive 
interactions~\cite{freezcomp}. As in these studies, the Helmholtz free energy
$A_{s}^{R}$ has been obtained by integrating with respect to density 
the equation of state of the hard-sphere solid given by Hall~\cite{hall},
starting from a density $\rho_{0}=0.736\, \rho_{cp}$, where 
$\rho_{cp}=\sqrt{2}\sigma^{-3}$ is the density at close packing,      
and setting the integration constant by the result 
$-A_{s}^{R}/\rho_{0}=5.91889$ determined via numerical simulation by Frenkel,
also reported in~\cite{foffi}.  
The radial distribution function of the hard-sphere solid $g_{s}^{R}(r)$ 
has been described by the parametrizations 
developed by Kincaid and Weis~\cite{kincaid} starting from 
Weis' Monte Carlo simulations~\cite{weis} 
and by Choi, Ree, and Ree~\cite{choi} on the basis 
of their own simulation results. These parametrizations share the same 
functional form, and differ only for the specific dependence 
of the parameters on the average density. In both cases, 
we used the data for the neighbor distance and coordination number 
of the face-centered cubic lattice reported in~\cite{hirschfelder}.
The integrals in Eqs.~(\ref{solid},\ref{mc}) were truncated at $r_{c}=8\sigma$,
and a long-range correction was added by setting $g_{s}^{R}(r)\equiv 1$ 
for $r>r_{c}$. Such a correction is actually irrelevant for the case discussed 
below.

\begin{figure}
\includegraphics[height=6cm, angle=0]{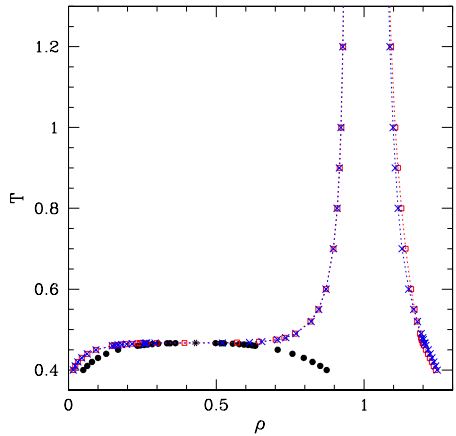}
\caption{Fluid-solid phase diagram of the HCYF for $z=5.6$. The fluid and 
solid phases have been described by HRT and perturbation theory respectively
(see Eqs.~(\protect\ref{solid},\ref{mc})). Full circles: HRT fluid-fluid coexistence
curve. The asterisk marks the position of the fluid-fluid critical point. 
Squares: freezing and melting lines obtained using the hard-sphere 
solid radial distribution function $g_{s}^{R}(r)$ by Kincaid 
and Weis~\protect\cite{kincaid}. Crosses: same as above using the 
$g_{s}^{R}(r)$ by Choi, Ree, and Ree~\protect\cite{choi}. Lines are a guide 
to the eye.}
\label{fig:freezing} 
\end{figure}

The fluid-solid phase boundary has been determined by equating the pressure 
and chemical potential of the solid given by Eqs.~(\ref{solid},\ref{mc}) 
to those of the fluid as predicted by the smooth cut-off HRT-OZ. 
As expected, we found 
that, as the inverse range $z$ is increased, the freezing line becomes wider, 
and for large enough $z$ it lies above the fluid-fluid coexistence curve. 
Figure~\ref{fig:freezing} shows the phase diagram for the case of marginal
stability, such that the freezing line is tangent to the fluid-fluid 
coexistence
curve at the critical point. This separates the stable fluid-fluid transition
regime from the metastable one. We have reported the results obtained using 
both Kincaid and Weis~\cite{kincaid} and Choi and Ree~\cite{choi} radial
distribution function $g_{s}^{R}(r)$. The two sets of results are nearly 
superimposed, save for some small deviations along the melting line. In both 
cases we obtain $z=5.6$ as the value of marginal stability. This result 
is very close to that given by SCOZA, $z=5.7$, via the same recipe used here 
to get Carnahan-Starling thermodynamics in the fluid phase,   
see Sec.~III, and is similar to the value $z=6$ determined by 
Hagen and Frenkel~\cite{hagen} via Gibbs Ensemble Monte Carlo simulations.
In the case of SCOZA, using the hard-sphere direct correlation 
function $c_{R}(r)$ given by Waisman~\cite{waisman} instead of fixing its 
range {\em a priori} gave $z=6.05$~\cite{foffi}, in closer agreement with 
the simulation result. It is likely that a similar trend would be shown 
by the smooth cut-off HRT-OZ with the same $c_{R}(r)$.      
  
\subsection{Equation of state and thermodynamics}

\begin{figure}
\includegraphics[height=6cm,angle=0]{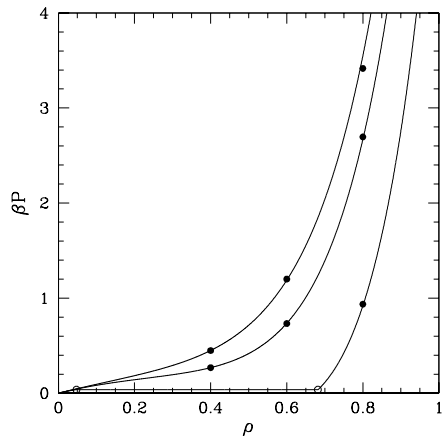}
\caption{Pressure divided by $k_BT$ 
as a function of density for $z=1.8$ at three temperatures:
$T=1$, $T=1.5$ and $T=2$ (from bottom to top). Line: HRT-OZ results. Full points: Molecular Dynamics
data form Ref. \cite{garnett}. The two open dots mark the boundaries of the HRT-OZ coexistence curve 
at the lowest temperature, characterized by a flat portion of the isotherm.  
}
\label{z18}
\end{figure}

\begin{figure}
\includegraphics[height=6cm,angle=0]{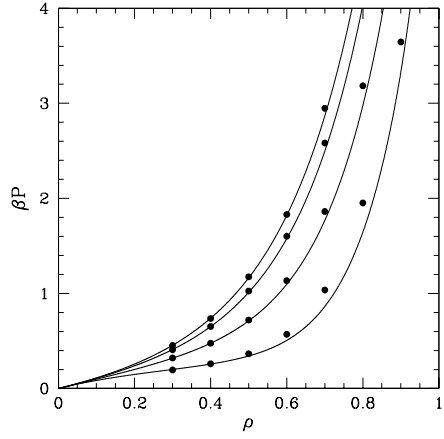}
\caption{Pressure divided by $k_BT$
as a function of density for $z=4$ at four temperatures: $T=0.7$,
$T=1$, $T=1.5$ and $T=2$ (from bottom to top). Line: HRT-OZ results. Full points: Molecular Dynamics
data form Ref. \cite{garnett}.
}
\label{z4}
\end{figure}

The equation of state of the HCYF can be easily obtained by integrating the HRT-OZ differential 
equation. The pressure divided by $k_BT$ 
is compared to Molecular Dynamics simulations
in Figs. \ref{z18} and \ref{z4} for $z=1.8$ and $z=4$ respectively. The agreement is very good  
showing that HRT-OZ is able to provide a consistent picture of thermodynamics throughout the 
phase diagram of the fluid. 

\begin{figure}
\includegraphics[height=6cm,angle=0]{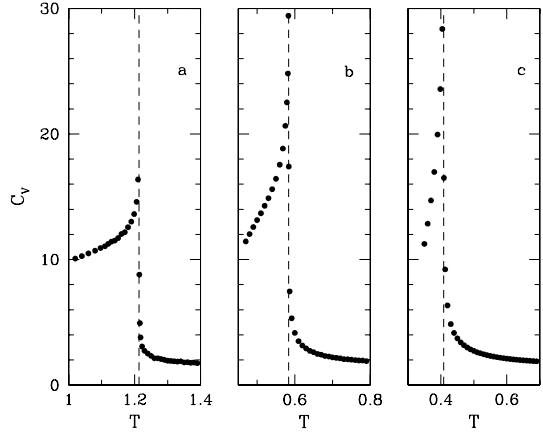}
\caption{Constant volume specific heat as a function of temperature at the critical density.
From left to right $z=1.8$, $z=4$ and $z=7$. The critical temperature is identified by 
a vertical dashed line.
}
\label{cvfig}
\end{figure}

Finally, in Fig. \ref{cvfig} we plot the specific heat as a function of temperature 
at the critical density for the three cases we have investigated $z=1.8$, $z=4$ and $z=7$.
The expected $\lambda$-line can be clearly seen in the HRT-OZ data. Note the different 
growth of $C_V$ at low temperature among the three cases.

\subsection{Correlation functions}

\begin{figure}
\includegraphics[height=6cm,angle=0]{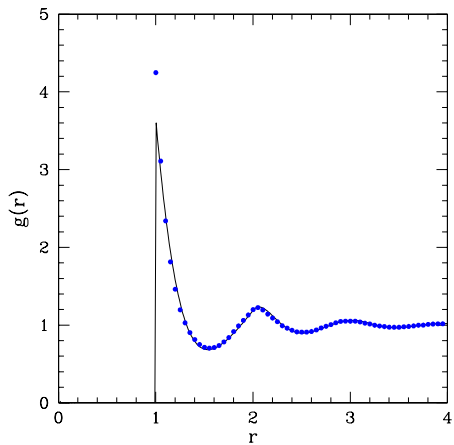}
\caption{Radial distribution function of the HCYF for $z=1.8$ at $T=1.5$ and $\rho=0.4$.
Full line: HRT-OZ results. Points: Monte Carlo simulation from \cite{scoza}. 
}
\label{corr1}
\end{figure}

\begin{figure}
\includegraphics[height=6cm,angle=0]{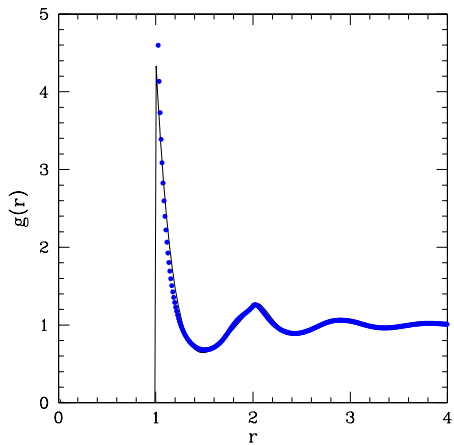}
\caption{Radial distribution function of the HCYF for $z=4$ at $T=0.5$ and $\rho=0.816$.
Full line: HRT-OZ results. Points: Monte Carlo simulation from \cite{pelli}.
}
\label{corr2}
\end{figure}

Although the Hierarchical Reference Theory has been especially devised for investigating the
thermodynamical properties of fluids, it also allows for the study of correlation functions
through the link between thermodynamics and correlations implied by the closure relation discussed
in Section III (\ref{comp},\ref{closc}). However, due to the MSA-like structure of the adopted 
scheme, we cannot expect an accuracy of the quality found for the thermodynamics. 
In Figs. \ref{corr1}-\ref{corr2} we compare the HRT-OZ results with Monte Carlo simulations 
for different values of $z$. The HRT-OZ result considerably underestimates the 
contact value of $g(r)$, especially at large $z$ or at low density, 
as a consequence of the ``linear" relationship between $c(r)$ and
the potential outside the core radius. 

As a side comment, we remark that the structure of correlations in the two phase region
is known to be analytically related to the correlation functions of the two pure phases: liquid (L)
and vapor (V). According to the theory of first order phase transitions \cite{fisher2}, the 
density correlations at coexistence are linear combinations of the density correlations 
for the two pure phases, leading to the following analytical structure 
of the radial distribution function at coexistence:
\beq
g(r)=
\left (\frac{\rho_L}{\rho}\right )^2 \frac{\rho-\rho_V}{\rho_L-\rho_V}\, [g_L(r)-1] + 
\left (\frac{\rho_V}{\rho}\right )^2 \frac{\rho_L-\rho}{\rho_L-\rho_V} \, [g_V(r)-1] +
\frac{\rho (\rho_L+\rho_V)-\rho_L\rho_V}{\rho^2}
\label{gcoex}
\eeq
This shows that in the two phase region, the radial distribution function $g(r)$ tends at 
large distances
to a non trivial limit, which differs from the expected uncorrelated result $g(r)\to 1$.
Such a behavior is in fact consistent with the divergence of the isothermal compressibility in the
two phase region, via the compressibility sum rule: 
$k_BT\rho\,\chi_{red}=1+\rho\,\int d\br \,(g(r)-1)$.  
Remarkably, the smooth cut-off HRT-OZ is able to capture this peculiar structure of the
two body correlation. In fact, it can be shown \cite{prl} that the asymptotic limit of $g(r)$
predicted by HRT-OZ  
precisely reproduces the last term of Eq. (\ref{gcoex}). However, further analysis shows 
that the closure adopted in this work is not able to describe the full structure of correlations 
in the two phase region expected on the basis of the exact result (\ref{gcoex}): a spurious 
long range decay in $g(r)$ appears inside the coexistence curve. 

\section{Conclusions}

In this work we have analyzed a simple Ornstein-Zernike 
closure to the Hierarchical Reference Theory of fluids in the smooth
cut-off formulation for hard core Yukawa fluids. Among the main advantages of this approach we 
can recall a consistent representation of the critical point, as customary in the HRT scheme, including
non classical critical exponents and scaling laws, together with a correct description of the first 
order phase boundary. Contrary to the usual sharp cut-off HRT-OZ, here the inverse compressibility shows the
expected discontinuity on the binodal which also allows to define the critical exponent $\gamma'$ describing
the divergence of the isothermal compressibility when approaching the critical point from the 
low temperature side. The expected long distance limit of two body correlations in the two phase 
region is also correctly reproduced by HRT-OZ. An estimate of the spinodal line can be extracted by 
studying the evolution of the inverse compressibility when density fluctuations of larger and larger 
wavelength are included. The thermodynamics of the HCYF agree well with simulations both near 
and far from the critical point: coexistence curves and equations of state can be predicted with high 
accuracy by HRT-OZ. 

Despite such a remarkable performance of HRT-OZ, there is still room for significant improvement.
Nowadays the most challenging applications of liquid state theory concern the so called complex fluids,
often characterized by extremely short range attractive interactions and/or more structured potentials
with both attractive and repulsive tails \cite{barrat}. The application of HRT to this class of
systems requires further work: the closure to the exact HRT hierarchy considered up to now is
clearly not adequate to describe complex fluids, for it assumes a linear relation between the
space dependence of the direct correlation function and the two body interaction. Moreover 
the rather artificial splitting between a short range ``reference" part of the potential $v_R(r)$ 
and a long range ``perturbation" $w(r)$, although customary in liquid state theory, limits the
applicability of the method to potentials where repulsive and attractive contribution can 
be easily identified. 
The HRT-OZ description of correlations, already for moderately short range
interactions, shown in Figs. \ref{corr1}-\ref{corr2}, is not fully satisfactory and should be improved. 
In Section III we pointed out a natural way to generalize the adopted closure, which paves the way to 
more elaborate, flexible and general approximation schemes able to combine the virtues of
the HRT approach and the accuracy of more sophisticated liquid state theories. We also mention 
a straightforward generalization of the smooth cut-off HRT method to binary mixtures, which is
potentially very useful in the framework of complex fluids. 

From a more fundamental point of view, few problems still remain open, and should be addressed
in future investigations. The HRT-OZ critical exponents are always affected by a systematic error
(which is of about $10\%$ for $\gamma$) clearly related to the adopted Ornstein-Zernike structure 
of correlations which implies $\eta=0$. This feature does not allow to apply the HRT-OZ 
approach in two dimensions. 
In order to go beyond this approximation, it is necessary to include, at least in some 
partial form, the effects of fluctuations on the momentum dependence of two body correlations:
an investigation of approximation schemes to the second equation of the HRT hierarchy (\ref{hrt1}) 
should be considered in the future. Hopefully, such a generalization will also provide a fully 
consistent picture of correlations in the two phase region. 

Finally, even within the present approximation scheme we did not address the subtle problem 
of the analytic structure of the free energy near the coexistence curve. According to the 
droplet model \cite{drop}, later confirmed by an exact result for the Ising model \cite{isakov},
an essential singularity is present at coexistence: the $n^{th}$ derivative of the free
energy with respect to the chemical potential should scale as $n!^{3/2}$ in three dimensions.
Some preliminary numerical evidence suggests that an essential singularity is indeed present also 
within HRT-OZ but determining the precise scaling requires some further analytical and numerical 
effort. 

We are grateful to Nigel Wilding, Dino Costa and Giuseppe Pellicane 
for providing simulation results of the radial distribution function 
of the HCYF.

\vskip 1cm
\centerline{\bf Appendix A: The HRT-OZ equations and their numerical solution}
\vskip 1cm
Here we discuss the formal structure of the smooth cut-off HRT-OZ equations (\ref{hrt0},\ref{comp})
specialized to the three dimensional ($d=3$) Yukawa fluid (\ref{yuk2},\ref{parz},\ref{park}).
We follow the notation of Ref. \cite{konior} (hereafter referred to as KJ)
for the solution of the MSA equations.
By substituting the form (\ref{yuk2}) into the evolution equation (\ref{hrt0}) we get: 
\begin{equation}
\frac{d A_t}{dt}=2\pi\rho^2\left [ C_1 \, \hat g_t(s) +C_2 \,\partial_s\hat g_t(s)\right ]_{s=z_2}
\label{hrt1b}
\end{equation}
Here $\hat g_t(s)$ is the Laplace transform of $r\,g_t(r)$:
\begin{equation}
\hat g_t(s)=\int_0^\infty r g_t(r) e^{-sr} \,dr
\label{lap}
\end{equation}
and the coefficients are defined as
\begin{eqnarray}
C_1 &=& \frac{e^{z}}{T}\,e^{-2t}\,(2\psi(t)-\dot \psi(t)) \\
C_2 &=& \frac{z}{T}\,e^{z}\,e^{-3t}\,\psi(t)
\label{pq}
\end{eqnarray}
where dots represent derivatives with respect to the evolution parameter $t$. 
According to Eq. (KJ-15)
\begin{equation}
\hat g_t(s) = \frac{s\tau(s)e^{-s}}{1-12\eta \,q(s)}
\label{gs}
\end{equation}
where $\eta$ is the packing fraction $\eta=\pi\rho/6$. 
The functions $\tau(s)$ and $q(s)$ are given by Eqs. (KJ-16,KJ-17,KJ-18) in
terms of ($z_1,z_2$) and the parameters $(a,b,\beta_1,\beta_2,d_1,d_2)$
which satisfy the equations
(KJ-10,KJ-11,KJ-12,KJ-13). This complicate set of six non-linear equations in six
unknowns allows to find the MSA parameters as a function of the input variables:
$(\rho,z_1,z_2,K_1,K_2)$. However, our specific problem is slightly different because
$K_1$ is actually unknown, depending on the consistency parameter $\lambda_t$. 
Instead, we would like to express $\hat g_t(s)$ in
terms of $(\rho,z,z_2,K_2)$ and the compressibility which appears into Eq. (\ref{comp}).
Luckily, the inverse compressibility is easily expressed in terms of the parameter $a$ by
\begin{equation}
\rho\,A_t^{''}=\rho\int d{\bf r} \,c_t(r) \,=1-a^2
\label{compa}
\end{equation}
where a prime represents differentiation with respect to $\rho$.
Therefore, in order to integrate the HRT-OZ differential equation (\ref{hrt1b}) we should solve, for
each $t$, the five equations (KJ-10,KJ-11,KJ-12) and the second of the pair of equations
(KJ-13) obtaining the five parameters $(b,\beta_1,\beta_2,d_1,d_2)$ in terms of
$(\rho,z_1,z_2,A_t^{''},K_2)$. Then, by substituting these expressions into Eq. (\ref{gs})
we can evaluate the right hand side of Eq. (\ref{hrt1b}) which becomes a partial differential 
equation (PDE) for the free energy density $A_t(\rho)$. Unfortunately, 
an analytic expression for the parameters $(b,\beta_1,\beta_2,d_1,d_2)$ is not available and 
then we decided to adopt a different procedure. 

As a first step it is convenient to change the parametrization of the MSA equations 
because, during the HRT evolution, the inverse range of the potential $z_2=ze^{-t}$ 
becomes extremely small giving rise to singularities in the parameters $(a,b,\beta_1,\beta_2,d_1,d_2)$. 
We first introduce $\tilde\beta_i$ and $\Delta_i$ defined by:
\begin{eqnarray}
\tilde\beta_i &=& \beta_i z_i^{-2} \\
d_i z_i^3 &=& -12\eta +\Delta_i
\end{eqnarray}
together with
\beq
k_2 = K_2\,z_2^{-4} 
\label{k2}
\eeq
which attains a finite limit for $t\to\infty$.
Then we note that the equations depend on the particular combination
$$
\gamma=b-12\eta \sum_i \tilde\beta_i
$$
which will be used in place of $b$. The new set of parameters is then given by
$(a,\gamma,\tilde\beta_1,\tilde\beta_2,\Delta_1,\Delta_2)$.
Actually only the renormalization of $\tilde\beta_2$ and $\Delta_2$ are necessary
but, for formal convenience, we adopted the same definition also for $\tilde\beta_1$ and $\Delta_1$.
The Laplace transform of the radial distribution function is written as
\begin{equation}
\hat g(s) = \frac{1}{s^2} + \frac{N(s)}{D(s)}
\label{laplace}
\end{equation}
where the ideal gas term has been written explicitly. The key quantities $N(s)$ and $D(s)$
are expressed in terms of the parameters by the following set of definitions:
\begin{eqnarray}
\label{alfa}
\Sigma &=& \sum_i \left [ -f_4(z_i)\,(-12\eta +\Delta_i) 
+f_1(z_i) \right ]\,\tilde\beta_i\frac{z_i}{z_i+s} \\
\Sigma_0 &=& -\,\sum_i \tilde\beta_i \,\frac{z_i}{z_i+s} \\
\Gamma &=& \sum_i \tilde\beta_i\Delta_i \frac{1}{z_i+s} \\
\tau(s)&=& a+(a+\gamma)\,s +s^2\,\Gamma -12\eta\,s\,\Sigma_0 \\
q(s) &=& f_5(-s)\,\tau(s) \\
P &=& \Sigma -\frac{a+\gamma}{24} -\frac{s}{24} \Gamma +\frac{1}{2}\eta \Sigma_0 +q(s) \\
N(s) &=& \tau(s) f_2(-s) +\Gamma -a -\gamma - s\,\Gamma +12\eta\Sigma_0  +12\eta\,P \\
D(s) &=& a+\gamma\,s -12\eta \,(s^2 \,P +s\Sigma_0)
\label{omega}
\end{eqnarray}
where we introduced the notation $f_n(x)$ for:
$$
f_n(x) \equiv x^{-n}\,\left [e^x -\sum_{j=0}^{n-1} \frac{x^j}{j!}\right ]
$$
which attains a finite limit as $x\to 0$.
The two equations KJ-12 and the single equation KJ-13 now become, respectively:
\begin{eqnarray}
\label{3}
z_1^3-\Delta_1 &=& 12\eta\,z_1^2\,\frac{N(z_1)}{D(z_1)} \\
\label{4}
z_2^3-\Delta_2 &=& 12\eta\,z_2^2\,\frac{N(z_2)}{D(z_2)} \\
\label{5}
k_2 \,z_2&=& \tilde\beta_2\,D(z_2)
\end{eqnarray}
while the first two equations KJ-10,KJ-11 are:
\begin{eqnarray}
\label{1}
\gamma&=&12\eta\left [ -\frac{a}{8}-\frac{\gamma}{6}-\sum_i \tilde\beta_i\Delta_i
f_3(z_i)+12\eta\sum_i\tilde\beta_i\,z_i\, f_4(z_i) + \sum_i\tilde\beta_i\,z_i\, f_1(z_i) \right ]\\
1-a &=& 12\eta\left [ -\frac{a}{3}-\frac{\gamma}{2}-\sum_i \tilde\beta_i\Delta_if_2(z_i)+
12\eta\sum_i\tilde\beta_i\,z_i\, f_3(z_i) + \sum_i\tilde\beta_i\,z_i\, e^{z_i} \right ]
\label{2}
\end{eqnarray}
By writing explicitly the ideal gas contribution to the radial distribution function
(\ref{laplace}) the evolution equation (\ref{hrt1b}) acquires an additional term
which precisely generates the mean field part of the free energy. This term 
can be absorbed into the definition of a modified free energy density
${\cal A}_t$:
\begin{eqnarray}
{\cal A}_t &=& A_t+A_{id}+\frac{\rho^2}{2}\left [ \tilde\phi(0)-\tilde\phi_t(0)\right] \nonumber \\
&=& A_t+A_{id}+ \frac{2\pi e^z}{T z^2} \rho^2 \,\psi(t)
\end{eqnarray}
where $\tilde\phi(k)$ denotes the Fourier transform of $\phi(r)=-\beta w(r)$.
Therefore, the evolution equation of the modified free energy density is
$$
\frac{\partial {\cal A}_t}{\partial t} 
= 2\pi\rho^2 \left [ C_1\frac{N(s)}{D(s)} + C_2 \frac{D(s)\partial_sN(s) 
-N(s)\partial_sD(s)}{D(s)^2}\right ]_{s=z_2}
$$
Now we define the new variable
\begin{equation}
u_t \equiv 2\pi\rho^2 \frac{e^z}{Tz^4}\, e^{2t}\,\left [ 
(2\psi(t)-\dot\psi(t))\,z_2^2\frac{N(s)}{D(s)} + 
\psi(t)\,z_2^3\frac{D(s)\partial_sN(s)-N(s)\partial_sD(s)}{D(s)^2} \,-\dot\psi(t)\right ]_{s=z_2} 
\label{defu2}
\end{equation}
which basically coincides with the right hand side of the evolution equation 
(\ref{hrt1b}), which in fact 
simply becomes
\begin{equation}
\frac{\partial{\cal A}_t}{\partial t} 
= z_2^2 \,\left [u_t + \frac{2\pi e^z}{T z^4}\rho^2 e^{2t}\dot\psi(t)\right ]
\label{evolb}
\end{equation}
Analogously, the compressibility equation (\ref{compa}) is expressed in terms
of the modified quantities as:
\begin{equation}
-\rho{\cal A}_t^{''} = a^2 - \rho\frac{4\pi e^z}{T z^4} \,z_2^2 \,e^{2t}\psi(t)
\label{compb}
\end{equation}
Combining Eqs. (\ref{evolb}) and (\ref{compb}) we finally get the evolution equation
for the bare dimensionless inverse compressibility $a^2$:
\begin{equation}
2a\dot a = -\rho z_2^2 u_t''
\label{last}
\end{equation}
In summary, we start from the definition (\ref{defu2}). We differentiate with respect to $t$,
which is also contained in $z_2$ and in the six parameters $(\gamma,\tilde\beta_i,\Delta_i,a)$.
The derivative of the first five parameters are calculated by differentiating the five equations
(\ref{3}-\ref{2}) 
and solving the linear problem in terms of $\dot a$. Finally, the derivative $\dot a$
is expressed in terms of $u_t''$ via Eq. (\ref{last}). This procedure provides a closed equation
for $u_t$ which must be coupled to the five algebraic non-linear equations
(\ref{3}-\ref{2}) and
to the definition (\ref{defu2}) in order to formally express the six parameters in terms of $u_t$.
An analytic solution of the set of equations 
(\ref{3}-\ref{2},\ref{defu2})
is clearly impossible, but we can implement a Raphson-Newton 
iterative scheme at each step in $t$, which in fact
converges very quickly.
The resulting evolution equation for the independent variable $u_t(\rho)$ has the general structure
\begin{equation}
\frac{\partial u_t}{\partial t} = g_1(u_t,t)+ g_2(u_t,t)\frac{\partial^2 u_t}{\partial \rho^2}
\label{pde} 
\end{equation}
which is a quasi-linear PDE in two dimensions and it is stable provided $g_2 \ge 0$.
This condition must be checked during the numerical integration and depends on the specific 
definition of the cut-off function $\psi(t)$. 
In this work we have considered two different choices for the function $\psi(t)$:
In the first part of the integration ($0<t<t^*$) we used
\begin{equation}
\psi(t)=\left ( 1+t/t_0\right )^{-2}
\label{first}
\end{equation}
where $t_0$ is a parameter. In the second part ($t>t^*$) we have chosen
\begin{equation}
\psi(t)=\left (\cosh t\right )^{-2}
\label{second}
\end{equation}
The transition point $t^*$ is obviously defined by the equation 
$1+t^*/t_0=\cosh t^*$. 

The initial condition at $t=0$ corresponds to a vanishing attractive tail 
$w_t(r)$ and can be obtained by substituting the analytic solution of 
the MSA equation for a hard sphere liquid, equivalent to the Percus-Yevick (PY) solution,
into the definition of $u_t$ (\ref{defu2}). This form of the reference correlations is 
however rather inaccurate at high density, where the PY solution shows appreciable deviations from 
the Verlet-Weis form \cite{vw}. A better choice is to parametrize the hard sphere direct
correlation function of the hard sphere liquid outside the core as a Yukawa tail. For convenience
we decided to fix the inverse range of this additional Yukawa equal to $z$, i.e. the inverse
range of the physical attractive part of the potential which defines the original problem.  
The amplitude of this additional contribution to the direct correlation function is fixed by 
requiring that the spatial integral of $c(r)$ is consistent with the Carnahan-Starling equation of 
state \cite{hansen} via the compressibility sum rule (\ref{compa}).
This choice is in fact equivalent to setting the initial value of the parameter $\lambda_t$ to 
a density dependent non vanishing limit. 
The numerical solution of the PDE also requires two boundary conditions at the minimum and 
maximum density: at $\rho=0$ $u_t$ vanishes due to the definition 
(\ref{defu2}) while at $\rho=\rho_{max}\sim 1$
the MSA approximation, corresponding to $\lambda_t=0$, provides a reasonable approximation. 
The PDE has been solved by a finite difference fully implicit predictor-corrector 
algorithm \cite{ames} 
with stability and accuracy checks during the numerical integration. The parameter $t_0$ in
the cut-off function has been chosen of order $1/z$ in order to preserve stability at all steps. 

\vskip 1cm
\centerline{\bf Appendix B: The asymptotic form of the HRT-OZ equations for a Yukawa potential}
\vskip 1cm
Let us first consider the simpler RPA based closure (\ref{closcrpa}) of the HRT-OZ evolution equation.
By substituting Eq. (\ref{closcrpa}) into (\ref{hrt0}) and by rescaling the momentum 
variable $q\equiv ke^t$, 
we see that for $t\to\infty$ the direct correlation function
$\tilde c_t(k)$ is evaluated in the $k\to 0$ limit and can be therefore expanded:
\beq
\tilde c_t(k) \sim \frac{\partial^2 A_t}{\partial \rho^2} -\,b\, k^2 
-\psi_\infty \,e^{-2t}\,\tilde\phi(ke^t)
\eeq
where $b$ attains a finite limit in the $t\to\infty$ also at the critical point and
is usually close to the range squared of the Yukawa tail while
$\psi_\infty$ is given by Eq. (\ref{asinto}). Notice that even in the $k\to 0$ limit, the
direct correlation function does depend on the precise form of the
two body interaction $\tilde\phi(q)$ which, in our case is just the three dimensional 
Fourier transform of 
the Yukawa potential divided by $k_BT$: $\tilde\phi(q)=4\pi\beta/(z^2+q^2)$. 
After some algebra, the evolution equation for the modified free energy (\ref{moda}) 
acquires the following form in three dimensions ($d=3$):
\beq
\frac{\partial {\cal A}_t}{\partial t} = 
\frac{e^{-3t}}{2}\int \frac{d^3q}{(2\pi)^d}\,
\frac{2\tilde \phi(q)- \tilde\phi^\prime(q)}{ (
-e^{2t}\,\frac{\partial^2{\cal A}_t}{\partial \rho^2}
+bq^2)\,\psi_\infty^{-1} +\tilde\phi(q)}
\label{hrt5}
\eeq
Let us introduce the parameter 
\beq
p=b z^4/(4\pi\beta\psi_\infty)
\label{prpa}
\eeq
measuring the ratio between
the curvature of the potential and of the direct correlation function (times $\psi_\infty$). 
Then we rescale the inverse compressibility via Eq. (\ref{defx}) leading to 
\begin{equation}
\frac{\partial {\cal A}_t}{\partial t} = e^{-3t}\,
\frac{z^3}{2\pi^2}\int_0^\infty dy\, \frac{y^2(1+2y^2)}{[1+y^2][1+(1+y^2)(x+py^2)]}
\end{equation}
The integral is well defined provided the denominator does not vanish in the
integration domain. Taking into account that $p>0$ we must have $x+1 > 0$ and either 
$p+x > 0$ or $p+x < 0$ and $x > p-2\sqrt{p}$. The latter condition may be satisfied only for $p <1$.
This integral is performed analytically by contour integration with the result (\ref{hrt6}),
valid for all $p$ and $x$ satisfying the above conditions.

It is now natural to rescale the physical quantities (free energy and density)
in order to get rid of the prefactors in Eq. (\ref{hrt6}):
\begin{eqnarray}
{\cal  A}_t(\rho) &=& \frac{z^3}{4\pi} \left [- p^{-1/2}\,\Psi_t(\varphi) + \frac{1-e^{-3t}}{3}
\right ]\nonumber \\
\rho -\rho_c &=& \varphi\,\left ( \frac{z^{5}}{(4\pi)^2\beta\sqrt{p}} \right )^{1/2}
\label{rescale}
\end{eqnarray}
In terms of the new variables $\Psi_t$ and $\varphi$ we simply get
\be
x &=& e^{2t}\, \frac{\partial^2\Psi_t}{\partial \varphi^2} \nonumber \\
\frac{\partial\Psi_t}{\partial t} &=& -e^{-3t}\,U(x)
\label{rescaleq}
\ee
where
\begin{equation}
U(x)\equiv\frac{2-p -\sqrt{p}\, \sqrt{1+x}} {\sqrt{p+x+2\sqrt{p}\, \sqrt{1+x}}}
\label{defu}
\end{equation}
which coincides with Eq. (\ref{hrt6b}). 
Differentiating the latter definition with respect to $t$, implicitly contained in $x(t,\varphi)$, 
we have
\begin{equation}
\frac{\partial U}{\partial t} =
\frac{dU}{dx}\left [ 2x - e^{-t} \frac{\partial^2 U }{\partial \varphi^2}
\right ]
\label{final}
\end{equation}
The equation is now written in quasi-linear form. We just have to formally obtain the coefficients
by inverting the definition (\ref{defu}) and evaluating $dU/dx$.

By defining the auxiliary quantity $y=\sqrt{p(1+x)}$ all the ingredients
appearing in (\ref{final}) are given by:
\begin{eqnarray}
x &=& \frac{y^2}{p}-1 \\
y &=& \frac{U^2(1-p)+(2-p)^2}{U p^{-1/2} \sqrt{U^2-p+4} + U^2 -p+2} \\
\frac{dU}{dx} &=& - \left [ 1+\frac{p}{2y}\right ] D^{-3} \\
D &=& \frac{4-p}{2U+\sqrt{p(U^2-p+4)}}
\end{eqnarray}

The asymptotic analysis of the evolution equation with the full MSA-based closure (\ref{closc}),
which includes the core condition, is more involved. 
By inspection of the equations for the MSA parameters (\ref{3}-\ref{2}), it is natural to guess that
$\gamma,\tilde\beta_1,\tilde\beta_2,\Delta_1,\Delta_2$ are all finite in that limit
Actually, outside the coexistence curve, $\Delta_2\to 0$, as shown by the
numerical solution of the equations. In particular, in the critical region, we may
assume that $\Delta_2\to 0$ and analyze the equations in the limit $z_2 \to 0$,
while $a/z_2$ and $\Delta_2/z_2$ are finite.
To leading order in $a$ and $z_2$, the equations KJ-10, KJ-11 and
the first of the KJ-12 just depend on $\tilde\beta_1$, $\Delta_1$ and
$\gamma_0=\gamma+12\eta\tilde\beta_1$.
Therefore, the two equations (\ref{4},\ref{5}) provide the
limiting values of $\tilde\beta_2$ and $\Delta_2/z_2$:
\begin{eqnarray}
\tilde\beta_2 &=& (12\eta)^{-1}\left [-(a/z_2+\gamma_0) +\sqrt{(a/z_2+\gamma_0)^2 + 24\eta k_2} 
\right ]\\
\Delta_2 &=& -12\eta\,\frac{
\tilde\beta_1\Delta_1/z_1 -\gamma_0 (1+\eta/2) +\tilde\beta_1(12\eta-\Delta_1)
f_4(z_1) +\tilde\beta_1 f_1(z_1)} { \sqrt{(a/z_2+\gamma_0)^2 + 24\eta k_2}}\,z_2
\end{eqnarray}
where $k_2$ has been defined in (\ref{k2}). 
When these expressions are substituted into the definition of $u_t$ (\ref{defu2})
the singular contribution to $u_t$ becomes
$$
u_t^{sing} \propto z_2\,\frac{\tilde\beta_2\gamma_0-4 k_2}{\sqrt{(a/z_2+\gamma_0)^2 + 24\eta k_2}}
$$
besides an additive constant.
Precisely at the critical point we expect that $a\sim z_2$,
while inside the coexistence curve the ratio $a/z_2$ will tend to the
value defined by the vanishing of the denominator: $(a/z_2+\gamma_0)^2 + 24\eta k_2=0$
(recall that $k_2 < 0$) leading to a divergence of $u_t^{sing}/z_2$ as $t\to\infty$
and to a finite limiting value of $\Delta_2$.
By comparing this expression with that obtained for a pure RPA (\ref{defu}), 
it is possible to identify the effective RPA parameters $p$ and $x$ as:
\begin{eqnarray}
\label{pmsa}
p &=& \frac{\gamma_0^2}{24\eta |k_2|} \\
1+x &=& \frac{a^2}{24\eta|k_2|\,z_2^2}
\end{eqnarray}
The requirement $0<p < 1$ then gives $0<\gamma_0^2 < 24\eta |k_2|$ which
should be verified in the region where phase transitions take place.

\newpage

\end{document}